\documentclass{article}
\usepackage{calor2kpro}
\begin{document}
\title{ 
  A LUMINOSITY SPECTROMETER FOR THE ZEUS EXPERIMENT AT HERA
  }
\author{
  Stathes D. Paganis \\
  {\em Nevis Laboratories, Columbia University, Irvington NY, 10533,
  USA} \\
  (On behalf of the ZEUS Collaboration)
  }
\maketitle
\baselineskip=11.6pt
\begin{abstract}
The HERA luminosity upgrade is expected to generate
two major problems in the current method of
luminosity determination which is based on
counting brehmsstrahlung photons: damage of the
calorimeter monitor due to high primary synchrotron radiation
and large multiple event (pile-up) corrections.
The luminosity spectrometer presented in this talk,
is a novel method that reduces the impact of
these problems in the luminosity measurement and is
expected to yield a total systematic uncertainty of 1.4\%.
The spectrometer counts brehmsstrahlung photon
conversions in the beam pipe exit window using
two small calorimeters (former ZEUS beam pipe
calorimeters) symmetrically placed
away from the synchrotron radiation plane.
The photon conversion rate is measured by counting
electron-positron (ep) coincidences in the calorimeters.
The ep acceptance is measured using a third calorimeter
(6 meter tagger) which tags the brehmsstrahlung electrons.
The electron-positron pair is separated by a small
dipole magnet.
\end{abstract}
\baselineskip=14pt
\section{Introduction}
ZEUS is one of the two colliding beam detectors operating at 
the HERA electron-proton collider at DESY, Germany. In 
September 2000, HERA completed an eight year running period which 
provided measurements of deep inelastic 
scattering (DIS) physics cross sections over a kinematic regime
that spans several orders of magnitude of the DIS parameters, 
Bjorken $x$ and photon virtuality $Q^2$ \cite{allen}. 
For the determination 
of the cross section $\sigma$ the accurate knowledge of the luminosity $L$
is required since $\sigma = R/L$ where $R$ is the corrected measured 
rate of a particular DIS process.
In ZEUS experiment the most accurate luminosity measurement achieved is 
1.1\%.

Currently HERA is undergoing a luminosity upgrade and is scheduled
to start the new physics run in August 2001. The goal is to increase 
the peak luminosity from $L\simeq 1.5\times 10^{31} cm^{-2}s^{-1}$ before 
the upgrade, to $L\simeq 7.5\times 10^{31} cm^{-2}s^{-1}$ after the upgrade.
The required accuracy in the luminosity measurement after the upgrade
is $1-2\%$. %imposed by the ZEUS physics goals.
The current ZEUS luminosity monitoring system has to be modified 
in order to meet the required accuracy because of two new major 
problems after the HERA luminosity upgrade:
\begin{enumerate}
\item Significant increase in the synchrotron radiation (SR).
\item Increase in the number of overlayed events (pile-up) for the physics 
process ($ep \rightarrow  ep\gamma$) used to measure luminosity, to 
$1.3-1.5$ per HERA bunch crossing.
\end{enumerate}
The luminosity spectrometer presented in these proceedings is a new 
method of measuring luminosity at ZEUS which solves the problem of 
SR and pile-up, meeting 
the requirements in the luminosity accuracy imposed by the 
ZEUS physics goals.

\section{The new ZEUS luminosity monitor}
\begin{center}
\begin{figure}[h]
  \vspace{6.0cm}
  \includegraphics{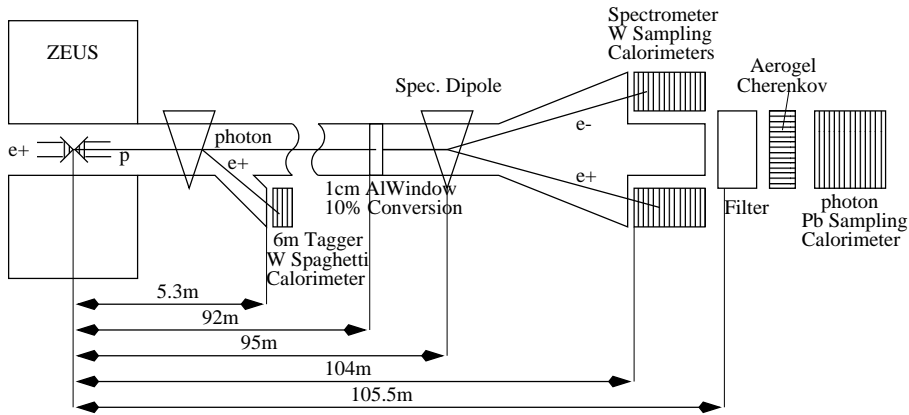}
  \caption{\it
    A schematic of the upgraded luminosity system of ZEUS (side
  view). A B-H photon produced at the IP converts 
  at the aluminum photon pipe exit window and the $e^+e^-$ coincidence 
  is recorded by two small W sampling calorimeters. The scattered 
  B-H positron is detected by the "e tagger", a W scintillating fiber
  calorimeter positioned 5.3~m away from the IP. Non-converted photons
  are detected by the photon Pb sampling calorimeter.
    \label{global} }
\end{figure}
\end{center}

\noindent
ZEUS has been succesfully measuring luminosity using the Bethe-Heitler
(B-H) $ep\rightarrow ep\gamma$ 
brehmsstrahlung process which is known theoretically within 
$0.5\%$. The photon is radiated with a very small angle with respect
to the e-p axis ($\leq 1mrad$) and in the past, the rate of these
photons was measured by a calorimeter positioned in the HERA tunnel 
107 meters away from the interaction point.
After the luminosity upgrade, $300-500~W$ of SR is expected to hit the
photon calorimeter at peak luminosity. Consequently one has to either
shield and upgrade the calorimeter or find some other method of 
measuring the luminosity. ZEUS decided to do both. The proposed and 
approved new luminosity system for ZEUS is shown in figure~\ref{global}.
In this figure the photon rate is measured by two independent 
methods. First the upgraded old method; 
a combination of two filters and two aerogel 
cerenkov counters is placed in front of the photon 
calorimeter in order to block the SR and correct for the photon 
energy loss in the filter. The calorimeter energy resolution 
depends on the accuracy of the energy loss correction performed. The 
photon pile-up problem though is not solved 
and certain corrections have to be performed in order to reduce the 
systematic error due to pile-up. 
The second and new method of measuring luminosity is the 
luminosity spectrometer which solves both the SR and 
pile-up problems by placing two small existing Tungsten 
scintillator calorimeters away from the SR plane. 
The photon rate is measured indirectly 
by counting $e^+e^-$ coincidences of photons converted
in the photon pipe exit window (about 10\%). 
The spectrometer has small acceptance in 
coincidences $\simeq3\%$ and no acceptance in low energy photons, 
so that the probability of a pile-up 
photon coincidence is small. The major contribution in the 
total systematic error of this method is due to the error in 
the knowledge of the acceptance. 

\section{The Luminosity Spectrometer}
\begin{center}
\begin{figure}[t]
  \vspace{9.0cm}
  \includegraphics{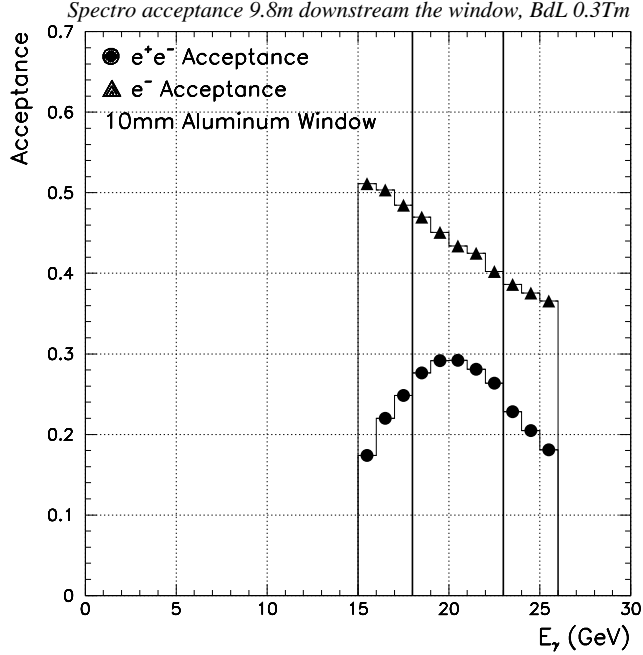}
  \caption{\it
Spectrometer acceptance as a function of the converted 
photon energy (conversion factor was omitted for simplicity).
    \label{acc1} }
\end{figure}
\end{center}

\begin{center}
\begin{figure}[t]
  \vspace{9.0cm}
  \includegraphics{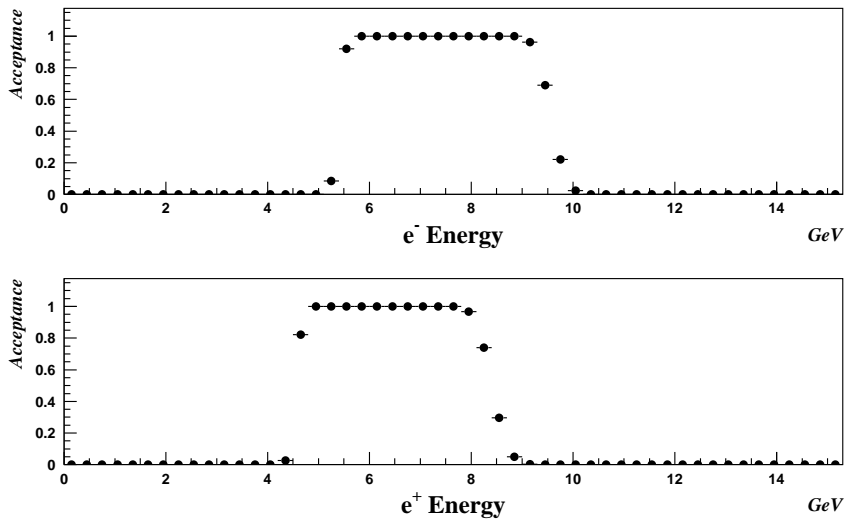}
  \caption{\it
Acceptance of the 6m tagger 
exit window for brehmsstrahlung electrons (top)
and positrons (bottom). The acceptance is $>99\%$ over the 
ranges $6-9 GeV$ for $e^-$ and $5-8 GeV$ for $e^+$.
    \label{6mt1} }
\end{figure}
\end{center}

The luminosity spectrometer consists of two well 
understood small ($12cm\times 12cm \times
24X_0$) Tungsten scintillator calorimeters 
which provide the energy and position of the incident electron,
and of a dipole magnet to separate the $e^+e^-$ pair. 
The calorimeter (former ZEUS Beam Pipe Calorimeter BPC) 
characteristics are listed in table~\ref{bpc}.
\begin{table}[t] 
  \centering
  \caption{\it 
    BPC characteristics.}
  \vskip 0.1 in
\begin{tabular}{|l|c|} \hline
 BPC specification  & BPC performance  \\
\hline \hline
 Depth & $24X_0$     \\
 Moliere radius     &  $13mm$    \\ 
 Energy resolution  &  $17\%/\sqrt{E}$ (stochastic term)    \\ 
 Energy scale calibration & $\pm0.5\%$     \\ 
 Energy uniformity &  $\pm0.5\%$    \\ 
 Linearity         &  $\leq 1\%$    \\ 
 Position resolution & $< 1mm$    \\ 
 Time resolution & $< 1ns$\\ \hline
\end{tabular}
\label{bpc}
\end{table}
The spectrometer acceptance $A$ is given by the formula:
\begin{equation}
A = A_{geom} \cdot A_{conv} \cdot A_z
\end{equation}
where $A_{geom}$ is the acceptance of the photon beam 
exit window (it is of order 90\%),
$A_{conv}$ is the conversion rate (for the specific exit 
window this is $A_{conv}\simeq 10 \%$), and $A_z$ describes the 
$z=E_e/E_\gamma$ dependent part of the acceptance. 
The spectrometer $ A_z$ part of the acceptance as a function of 
the converted photon energy is 
shown in figure~\ref{acc1}. The $e^+e^-$ coincidence 
acceptance is relatively uniform in the 18-23 GeV photon 
energy window. All of these accepted coincidences are coming 
from symmetric conversions i.e. they have 
$z=0.5 \pm 0.2$.
In the spectrometer method the 
acceptance is actually measured using a new small Tungsten 
scintillating fiber calorimeter, the 6-meter tagger (6mt) 
placed 5.3 meters away from the IP. The 6mt
has 100\% acceptance in brehmsstrahlung electrons 
in an energy window where the spectrometer 
acceptance is maximized (figure~\ref{6mt1}).
A coincidence between the 6mt and the
spectrometer should give a total energy equal to the 
electron beam energy: $E_e+E_\gamma=27.5~GeV$. 

The expected coincidence rates are high
thus providing an online statistically significant luminosity
measurement. The converted B-H photon Gaussian profile on 
the exit window can be reconstructed within a few seconds.
The $e^+e^-$ origin Y coordinate $Y_{hit}$ is 
approximated as the energy-weighted position of the two electrons
as measured in the BPC's:
\vspace{0.2cm}
\begin{equation}
Y_{hit} = \frac{E^+\cdot BPC^+ + Y^-\cdot BPC^-}{E^+ + E^-}
\end{equation}
where $E^+, BPC^+$ and $E^-, BPC^-$ the energies and positions of 
the $e^+$ and $e^-$ as obtained from the BPC's.
The resolution obtained is around 5~mm
with a 2~mm accuracy in the mean of a gaussian assumed profile. 
The photon profile will provide to HERA beam monitoring information.

A list of the most significant systematic errors of the spectrometer
method is given in
table~\ref{summary} \cite{proposal}.
\begin{table}[t] 
  \centering
  \caption{ \it 
    Systematic Error in luminosity measurement.
    }
  \vskip 0.1 in
\begin{tabular}{|l|c|} \hline
 Error Type  & $L=7\cdot 10^{31}cm^{-2}s^{-1}$\\
\hline \hline
 Multiple event correction & $\leq 0.5\%$ \\
 egas bgnd subtraction     & $\leq 0.5\%$ \\
 Total Acceptance error    & $\leq 1.0\%$ \\
 Energy Scale errors       & $\leq 0.5\%$ \\
 Cross-section Calculation & $\leq 0.5\%$ \\
% Counting Errors           &  -    &    -         \\ \hline
% Thermal $\gamma$ background & -   &    -         \\ \hline
% p-beam background         &   -   &    -         \\ \hline
\hline
 Total systematic error & $\leq 1.4\%$ \\ \hline
\end{tabular}
\label{summary}
\end{table}
The counting, thermal $\gamma$ background and proton-beam background 
errors are currently expected to be relatively low so they are 
not taken into account. The theoretical calculation of the cross
section error was taken as $0.5\%$. The total systematic error 
is calculated by summing all errors in quadrature.
The total systematic error is expected to be below $1.4\%$. 

\section{Conclusions}

The luminosity spectrometer method for measuring 
luminosity in the ZEUS experiment 
at HERA was presented.
ZEUS is measuring luminosity by counting photons 
produced by the Bethe-Heitler $ep\rightarrow ep\gamma$ process.
A 10\% of these photons are converted in a thin exit window.
The spectrometer measures the converted photon rate
by counting $e^+,e^-$ coincidences 
in two small well understood calorimeters. The calorimeters 
are away from the SR plane and pile-up is a secondary effect.
The spectrometer acceptance is measured using an independent 
device, the 6m e-tagger. The tagger has almost 100\% acceptance 
in B-H electrons in an energy window where the spectrometer 
acceptance is maximized. The electron and photon energy sum 
equals the electron beam energy ($27.5~GeV$).
Detailed calculations show that the luminosity spectrometer
can measure luminosity with an accuracy better 
than 1.4\%~\cite{proposal}.
%The ZEUS Luminosity measurement after the HERA
%upgrade has two difficulties 
%synchrotron radiation
%and pile-up (two Bremss. photons per bunch crossing).
%The old method can still be used after appropriate 
%modification of the current setup: measure directly 
%the photons using filters to block the SR and counters 
%to correct for energy loss in the filters. The challenge 
%for this method is the pile-up correction and energy scale
%calibration.


\begin{thebibliography}{99}
\bibitem{allen}
H. Abramowicz, A. Caldwell, Rev. Mod. Phys. {\bf 71}:1275-1410, (1999).
\bibitem{proposal}
A. Caldwell, S. Paganis, R. Sacchi, F. Sciulli, 
{\it A Luminosity Spectrometer for ZEUS}, Internal ZEUS document, (1999).


\end{thebibliography}
\end{document}